\documentclass[pdflatex,sn-mathphys-num]{sn-jnl}

\usepackage{graphicx}%
\usepackage{multirow}%
\usepackage{amsmath,amssymb,amsfonts}%
\usepackage{amsthm}%
\usepackage{mathrsfs}%
\usepackage[utf8]{inputenc}
\usepackage[title]{appendix}%
\usepackage{textcomp}%
\usepackage{manyfoot}%
\usepackage{booktabs}%
\usepackage{algorithm}%
\usepackage{algorithmicx}%
\usepackage{algpseudocode}%
\usepackage{listings}%
\usepackage[english]{babel}
\usepackage{amsmath} 
\usepackage{amsthm} 
\usepackage{amssymb}
\usepackage{amsfonts}
\usepackage{multirow}
\usepackage{multicol}
\usepackage{pdflscape}
\usepackage{multirow}
\usepackage{graphicx}
\usepackage{ulem} 
\usepackage{hyperref}
\usepackage{url}
\usepackage{enumerate}
\usepackage{graphicx}
\usepackage{epstopdf} 
\usepackage{tikz}
\usetikzlibrary{arrows.meta,positioning}
\usepackage{geometry}
\theoremstyle{thmstyleone}%
\theoremstyle{thmstyletwo}%

\theoremstyle{thmstylethree}%

\begin{document}

\title[Article Title]{Modeling of Pneumococcal and Respiratory
 Syncytial Virus Pneumonia: An Epidemiological
 Review, with Statistical Inference}
\author*[1]{\fnm{Rupchand } \sur{Sutradhar}}\email{rupchandsutradhar@gmail.com}
\equalcont{These authors contributed equally to this work.}

\author[1]{\fnm{Anuj } \sur{Mishra}}\email{anuj.mj@gmail.com}
\equalcont{These authors contributed equally to this work.}

\author[1,2]{\fnm{Malay } \sur{Banerjee}}\email{malayb@iitk.ac.in}
\equalcont{These authors contributed equally to this work.}

\author[1,2]{\fnm{Subhra Sankar } \sur{Dhar}}\email{subhra@iitk.ac.in}
\equalcont{These authors contributed equally to this work.}

\affil[1]{\orgdiv{Mathematics and Statistics}, \orgname{Indian Institute of Technology
 Kanpur}, \orgaddress{ \city{Kanpur}, \postcode{208016}, \state{Uttar Pradesh}, \country{India}}}

\affil[2]{\orgdiv{National Disease Modelling Consortium}, \orgname{Indian Institute of Technology Bombay}, \orgaddress{ \city{Mumbai}, \postcode{400076}, \state{Maharashtra}, \country{India}}}
 

\abstract{Infectious diseases continue to pose significant public health challenges worldwide, requiring effective prevention and control strategies to mitigate their negative impact. Infectious diseases can be broadly classified into two groups: vaccine-preventable diseases (\textit{e.g.}, measles, polio, influenza, hepatitis B, pneumonia) and vaccine-non-preventable diseases (\textit{e.g.}, HIV/AIDS). Vaccine-preventable disease models are one of the essential tools for understanding infectious disease dynamics, evaluating intervention strategies, and guiding public health policies. In this review article, we explore the recent advancements in modeling two particular vaccine-preventable infectious diseases. Here, we consider both deterministic and stochastic models to comprehensively capture the complexity of disease transmission, vaccine efficacy, and population-level immunity. We highlight the application of these models to the infectious diseases, namely, bacterial and viral pneumonia caused by the bacteria  \textit{Streptococcus pneumoniae} (\textit{S. pneumoniae}) and the respiratory syncytial virus (RSV). Pneumonia carry a substantial global burden, where modeling has played a crucial role in assessing vaccine impacts and optimizing immunization strategies to minimize the disease burden. By synthesizing recent methodologies and findings, this review provides valuable insights for future research and policy decisions aimed at improving vaccine-preventable disease control for pneumonia caused by \textit{S. pneumoniae} and RSV.}


\maketitle

\section{Introduction}\label{sec1}
Pneumonia is one of the most clinically significant vaccine-preventable disease due to its high incidence worldwide. It is a form of acute respiratory infection that is most commonly caused by viral and bacterial pathogens and can range in severity from mild to life-threatening disease across all age groups. Despite being preventable and treatable, pneumonia continue to pose  a substantial  mortality burden in children, causing approximately 808,000 deaths in children under five years of age in 2017, which accounts for about 15\% of all under-five mortality \cite{WHO}. This infection is generally transmitted through the direct contact with the infected individuals. 

\subsection{Background: Vaccine Preventable Diseases}
Vaccine-preventable diseases (VPDs),  such as pneumonia, measles, influenza, hepatitis A, hepatitis B, human papilloma virus (HPV), etc. infection remain a leading cause of morbidity and mortality throughout the world, particularly among vulnerable populations including children under five years of age, the elderly, and immunocompromised individuals and those with chronic illnesses \cite{WHO,trombetta2022influenza,parums2024review,2015_liu_global,al2024global,stanley2014human}. Pneumonia is an inflammation of the air sacs (alveoli) in the human lungs, which become filled with fluid or pus (a thick, yellowish or greenish fluid). It is a treatable respiratory infection, typically managed with antibiotics, antiviral, or antifungal medications depending on the underlying cause, and supportive care. Sometimes it needs hospitalization and other special medical support.   This pulmonary infection can be caused by a variety of pathogens, including bacteria, viruses, fungi, and parasites. Among these, \textit{Streptococcus pneumoniae }(\textit{S. pneumoniae}) bacteria and Respiratory syncytial virus (RSV) are two of the most significant contributors to severe lower respiratory tract infections \cite{2009_brien_burden,2010_nair_global}.

\subsubsection{\textit{S. pneumoniae} and Its Associated Diseases}

\textit{S. pneumoniae} is a gram-positive, lancet-shaped diplococcus and one of the most common bacterial pathogens responsible for bacterial pneumonia, in particular community-acquired pneumonia (CAP) \cite{feldman2016role}. This bacteria colonize the nasopharynx and can invade the lower respiratory tract, leading to a range of diseases, including otitis media (ear infection), sinusitis (sinus infection), meningitis (brain and spinal cord infection), and bacteremia (bloodstream infection), in addition to pneumonia \cite{Brooks2018,weiser2018streptococcus,shak2013influence}. This bacterium is encapsulated, with more than 90 known serotypes \cite{2009_brien_burden}, and the polysaccharide capsule is a major virulence factor that helps it evade phagocytosis \cite{2004_bogaert_streptococcus,2009_van_pathogenesis}. 
Vaccines such as the pneumococcal conjugate vaccine (PCV13) and the pneumococcal polysaccharide vaccine (PPSV23) are key public health tools in preventing this infection \cite{Le2024}.  Globally, pneumococcal disease imposes a heavy burden, especially in low- and middle-income countries where vaccine coverage and healthcare infrastructure may be limited \cite{2000_wp_pneumococcal,2011_weinberger_serotype}. Invasive pneumococcal disease (IPD) predominantly affects children under the age of two years and adults over 65 years of age, manifesting clinically as pneumonia, meningitis, and sepsis \cite{2011_klugman_contribution,2018_wahl_burden}.

\subsubsection{RSV and Its Associated Diseases}

RSV is an negative-sense, enveloped, single-stranded RNA virus belong to  the \textit{Paramyxoviridae} family and is a leading cause of bronchiolitis (bronchioles infection) and pneumonia, particularly in infants and young children under two years of age \cite{who_RSV}.  However, most individuals acquire the infection by the age of two. RSV is classified into two main subtypes (RSV-A and RSV-B) based on differences in their surface proteins, especially  G glycoprotein responsible for virus attachment to host cells mainly epithelial cells lining the upper and lower respiratory tract \cite{nuttens2024differences}. RSV infection is highly seasonal, causing recurrent outbreaks that exert significant pressure on healthcare systems worldwide \cite{2009_cb_burden}.  This infection can spread mainly through respiratory droplets and direct contact with contaminated surfaces including  doorknobs, toys and other objects that are frequently used everyday \cite{CDC_RSV_Spreads_2025}. Note that  this infection causes inflammation and obstruction of the small airways, called bronchioles, often leading to the following symptoms: wheezing, coughing, and respiratory distress. It is important to note that while most infections are self-limiting and are cured by immune system, several cases may require hospitalization and clinical support, particularly in premature infants, immunocompromised person, and those having congenital heart or chronic lung infection \cite{Shi2017,chatterjee2021current,kaler2023respiratory}. 

In practical, the prevention strategies that are widely applied include health hygiene, avoiding exposure during the peak season, and the used of monoclonal antibody for high-risk groups. The long-used of monoclonal antibodies are also considered as a promising preventive options for RSV infection \cite{cleo2025}.  Although this infection has been studied  for several decades, the absence of a licensed vaccine is one of the reasons for the difficulties arising from its intricate immunology and pathogenesis \cite{2005_falsey_respiratory,2017_chan_burden}. Therefore, it is an urgent need  to have  more effective and targeted interventions for this viral infection, focusing vaccination and   emphasizing treatment strategies that minimize the use of antibiotic.
 According to the reports of World Health Organization (WHO) and other diseases control and prevention monitoring platforms, pneumonia is responsible  for approximately 15\% of all deaths in children under the age of  five\cite{2011_nair_global,troeger2018estimates}. In addition, the recent COVID-19 pandemic underscores the critical importance in understanding the respiratory infections and implementing effective control and prevention measures like knowledge is required for guiding the public health interventions, protecting the vulnerable populations, specifically infants and children, developing therapies and vaccines, and finally reducing the burden on healthcare systems globally \cite{2020_devakumar_impact,2024_zhou_impact}.

\subsection{Literature Review of Mathematical Modeling for Bacterial and Viral Pneumonia}
In order to understand the spread  and the dynamics of the infection, and control of infectious diseases, mathematical modeling plays an extensive roles. Note that the pathogens like \textit{S. pneumoniae} and RSV have complex biological  and immune-epidemiological features that need careful analysis and through investigations \cite{Mumbu2024}. Note that mathematical models also contribute  by integrating different types of data, such as immunological, clinical and genetic information to simulate disease dynamics. Mathematical models have extensive to predict outbreaks, and evaluate effective interventions. Several approaches are applied, such as clinical data related population models \cite{keeling2008modeling}, immunological data inform within-host models \cite{perelson1999mathematical}, and genetic data guide pathogen evolution studies \cite{luksza2014predictive}. In the literature, researchers have used a range of modeling approaches, such as deterministic models, age-stratified models, age-structured models,  vaccination epidemic models, and stochastic models. These models are primarily proposed to study different aspects of disease transmission and facilitate optimal decision-making,  and developed further as per requirements \cite{Teklu2024}. 
In this context, the classical epidemic models, SIR (Susceptible-Infectious-Recovered) and its extensions (SIRS, SEIR, SEIRS), are foundation in epidemic modeling. 

To reflect the complexity of the disease transmission in a better way,  numerous mathematical models have been proposed and developed incorporating various  epidemiological and biological factors, including  waning and natural immunity \cite{2024_purohit_novel,reicherz2022waning}, latency periods \cite{2008_Temime_1,2013_Otieno}, age-structure and demographic changes \cite{2008_Temime_1}, seasonality \cite{2008_Temime_1,moore2014modelling,leecaster2011modeling,numminen2015climate,domenech2019unraveling,weinberger2014seasonal}, spatial and spatio-temporal heterogeneity \cite{wangdi2021spatio,liang2025associations,degif2025spatial},  the role of asymptomatic carriers \cite{2013_Otieno}, and a serious of model dealing with control strategies including vaccination and optimal interventions, etc \cite{2021_swai_optimal}.     
As the children under five years of age and adults over 65 are at high risk of pneumonia, age-stratified models are helpful for accurately reflecting the epidemiology scenarios. These models often divide the whole population into  several age cohorts, capturing differences in susceptibility of the infection, ability to exposes, and disease outcomes. In this context, the recent studies  developed several models focusing specifically on these high-risk groups (children under five years of age and adults over 65), and  enable targeted intervention assessments and resource allocation \cite{Chukwu2024}. 
Based  on these advancements, a systematic review of mathematical models on bacterial and viral pneumonia  is  necessary. While  numerous models have been introduced to study disease dynamics from different perspectives, a comprehensive synthesis of their modeling frameworks can offer a  deeper insights into their shared and unique transmission mechanisms, epidemiological complexities, and  intervention impacts. It is expected that such a review will not only  help to track the evolution of both deterministic and stochastic modeling approaches, but also highlight methodological innovations, and determine the key knowledge gaps that indicate further investigation. Moreover, examining how the available models in the literature have been applied to evaluate the vaccination strategies, the waning immunity and  co-infections, etc. will enhance the evidence base for public health decision-making and guide the design of control strategies and  more effective prevention.

In 2015, Ndewla \textit{et al.} \cite{ndelwa2015mathematical} proposed a mathematical model to quantify the treatment and the screening strategies for pneumonia. In this context, Tilahun \textit{et al.} \cite{2017_Tilahun} formulated a five-compartmental model classifying the entire population into the following groups: Susceptible, Vaccinated, Carrier, Infected, and Recovered yielding the  well-known SVCIR frameworks while incorporating optimal control measures and cost-effectiveness analysis. In a similar manner, Kizito \textit{et al.} \cite{2018_kizito_mathematical} introduced a SCIR (Susceptible-Carrier-Infected-Recovered) model to study the transmission of bacterial pneumonia, concentrating on the synergetic effects of treatment and vaccination. In the same pipeline, Opara \textit{et al.} \cite{zephaniah2020mathematical} proposed a model in 2020  based on the SVEIR (Susceptible-Vaccinated-Exposed-Infected-Recovered) architecture to explore the transmission dynamics of pneumonia in the presence of a preventive and effective vaccine. It is also noticed that Jilan  \textit{et al.} \cite{alya2022mathematical} constructed a model and addressed the spread of pneumococcal (bacterial) pneumonia, considering both vaccination and hospital care, distinguishing between treated and untreated individuals. Furthermore, Mochan \textit{et al.} \cite{mochan2014mathematical} presented a more simplified system to model the intra-host immune response to bacterial pneumonia by taking into account the  diverse immune profiles that are observed in various murine strains.
No doubtfully, besides  these individual contributions, a substantial body of literature has focused on the development of mathematical models that explicitly include  the effects of vaccination to control pneumonia \cite{2017_Tilahun, 2025_ALDILA_100394, 2022_kotola_mathematical, 2012_greenhalgh_mathematical, 2021_swai_optimal}.

Due to the availability of these versatile and  robust models, a set of values of basic reproduction number ($R_0$)  for these infectious diseases is reported. Note that these variations in $R_0$ help people capture diverse epidemiological settings, quantify the impacts of parameters, evaluate transmission potential under different modeling assumptions,  identify the most influential factors driving disease spread, measure the sensitivity of inputs and evaluate the impacts of intervention strategies. In the end, these differences offer  a deeper understanding of pneumonia dynamics and support to formulate the robust, evidence-based control strategies and prevention policies.
 Besides the development of mathematical models for pneumonia caused by \textit{S. pneumoniae}, extensive research has also been devoted to modeling RSV infection. Various studies have introduced different modeling components, including separate compartments for RSV groups A and B \cite{kombe2019model}, multiple types of infectious compartments \cite{hodgson2020evaluating,kinyanjui2020model,pan2017predicting}, waning of partial immunity to reinfection among susceptible individuals \cite{mahikul2019modeling,kinyanjui2020model,pan2017predicting}, and the use of multiple nested dynamic transmission frameworks \cite{arenas2008existence}. Recently, the US Centers for Disease Control and Prevention developed a static model to estimate the burden of medically attended RSV infections under several intervention strategies \cite{rainisch2020estimating}. However, while static models are effective in estimating the direct effects of interventions, they are limited in capturing indirect or herd immunity effects, which are often significant in infectious disease dynamics \cite{pitman2012dynamic}. Overall, several other studies addressing both these disease have also been conducted, which will be discussed in detail in the subsequent sections.

 \subsection{Epidemiological context}

 Mathematical models of these diseases transmission provide key epidemiological indicators, including the basic and effective reproduction numbers, force of infection, incidence, and prevalence, etc \cite{anderson1991infectious,diekmann2013mathematical}. These models are used to estimate latent and infectious periods, transmission rates, recovery and waning immunity, and age-specific susceptibility and hospitalization risks, which are essential for understanding disease spread in high-risk groups. Vaccination models further quantify vaccine efficacy, coverage requirements, and herd immunity thresholds, while advanced models capture co-infection dynamics. Together, these parameters enhance understanding of these diseases epidemiology and guide effective, evidence-based public health interventions \cite{oidtman2025modelling}.
 
\begin{figure}[h]
    \centering
    \includegraphics[width=15cm, height=10cm]{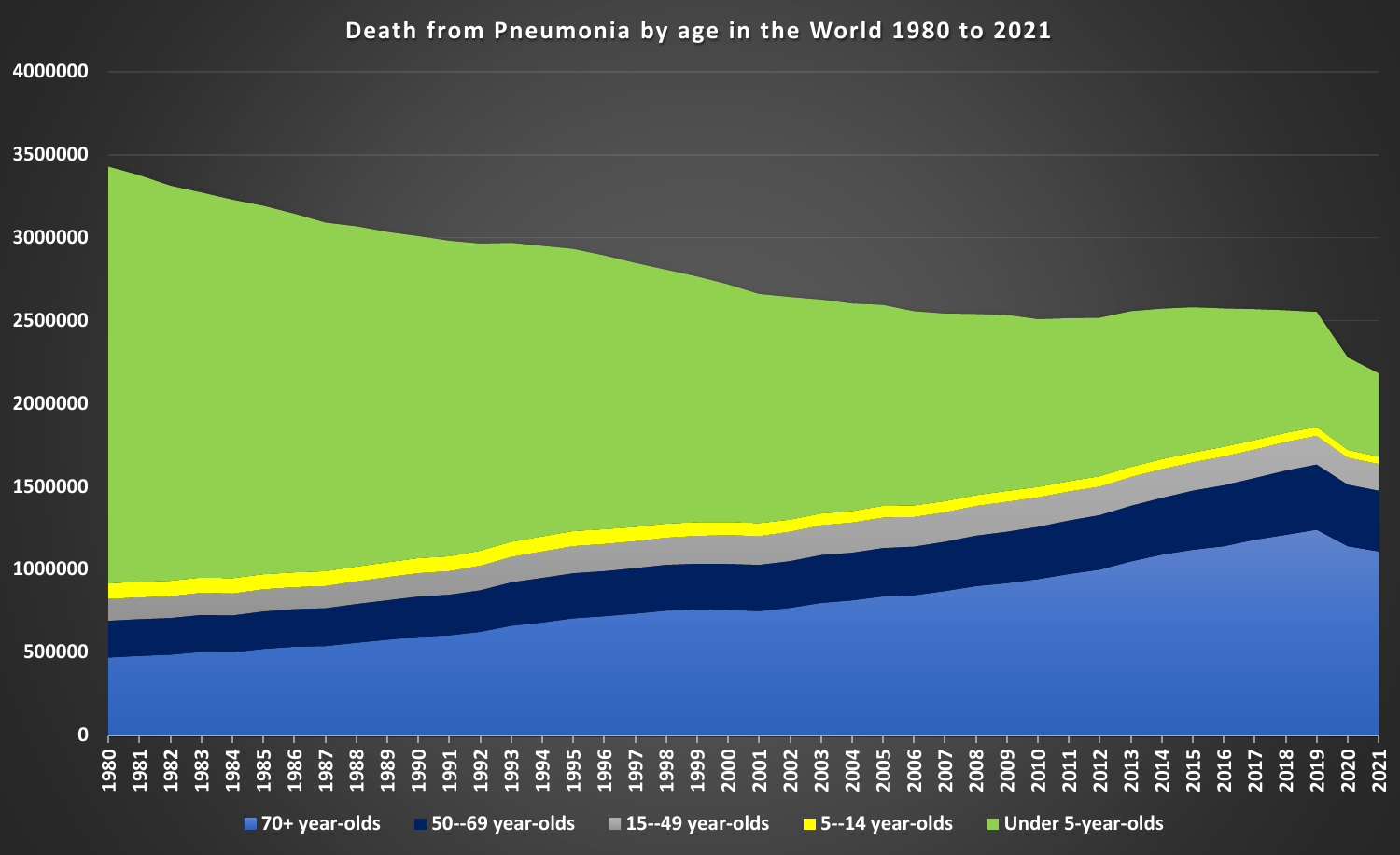}
    \caption{The chart shows the estimated number of annual death  from pneumonia in each age group worldwide. The data are collected from \cite{worlddata}.}
    \label{fig:annual death}
\end{figure}

According to WHO, pneumonia remains one of the leading causes of death among mainly children under five years of age and continues to pose global health issue, in spite of having significant reductions achieved over recent decades  due to the improvements in immunization, treatment protocol, sanitation, and healthcare access \cite{worlddata,troeger2018estimates}. Note that as illustrated in Figure \ref{fig:annual death}, pneumonia mortality rates shows distinct patterns in different age groups. One can see that there is a marked decline in mortality rate among young children since 1980. The  primary reason is the implementation of   widespread pneumococcal conjugate vaccine (PCV) and the improvement in clinical management. Note that in order to reproduce  the Figure \ref{fig:annual death}, the annual mortality data of pneumonia for each age group calssified as under 5 years, 5–14 years, 15–49 years, 50–69 years, and 70+ years are taken from the open-source database 'Our World in Data'
 \cite{worlddata}. These data are first cleaned, compiled and organized to make sure consistency across age groups. This  allows accurate visualization and enables comparison of pneumonia-related deaths over time worldwide. The trend of the data shows that there is a rapid decrease in the mortality rate in pneumonia in 2019. On the other hand, older adults experience continuously the highest pneumonia-related death. This reflects the age-associated immune decline, comorbidities or coexisting diseases, and reduced vaccine efficacy \cite{torres2018burden,worlddata}. Nevertheless, the overall percentage of pneumonia-related deaths irrespective of age remains substantial; this  underscores the persistent public health challenge posed by this disease. So, more efforts toward the prevention of pneumonia, early diagnosis facility, and effective treatment are really necessary to further reduce pneumonia-related mortality and improve health outcomes.

Beyond its clinical significance, pneumonia also shows complex epidemiological interactions, and  exhibit seasonal and demographic variability in wider window. To understand  these intricate transmission patterns and disease dynamics,  mathematical and statistical modeling frameworks are required as by the help of these models, one can quantify infection parameters, predict future disease trends and assess intervention effectiveness. Therefore, combining the epidemiological modeling with the robust statistical inference is vital to improve the understanding of pneumonia transmission in population level and inform the evidence-based public health strategies for better diagnosis. 
\color{black}

\color{black}

\subsection{Scope of this review}

Unlike a systematic review \cite{munn2018systematic}, this review does not focus on a particular research question. The primary objective of this review is to summarize existing pneumonia transmission models to inform future dynamic modeling studies. Specifically, we discuss the current state of modeling for pneumococcal and RSV pneumonia, with an emphasis on three specific types of models:

\begin{enumerate}[(i)]
    \item \textbf{Epidemic models}: Epidemic models are mathematical frameworks used to describe how infectious diseases spread within a population over time. For example, common epidemic models mainly include the SIR model, SIS model, and SEIR model and their straightforward extensions. They help researchers and policymakers understand transmission dynamics, predict outbreaks, and design effective control strategies such as vaccination or quarantine. For this study, we do not consider the effects of randomness and vaccination in the epidemic model. Epidemic models include compartmental, age-structured, network-based, and meta-population frameworks that capture only the transmission dynamics of disease spread.
    \item \textbf{Stochastic models}: Stochastic models capture the randomness in disease transmission, especially in small populations. They use probabilities to describe infection and recovery events, providing realistic predictions of outbreak variability and extinction chances.
    \item \textbf{Vaccination models}: Vaccination models study how immunization affects disease spread and control. These models explicitly incorporate vaccination into disease dynamics and assess its impact through various strategies, including evaluating direct and indirect effects, serotype replacement, and integration with non-pharmaceutical interventions to optimize disease control and prevention.
\end{enumerate}
\noindent
By synthesizing advances across epidemiology, immunology, computational biology, and public health, mathematical modeling has emerged as a critical tool for understanding and managing pneumonia resulting from viral infections such as RSV and bacterial infections caused by \textit{S. pneumoniae}. These models are evaluated based on their structural assumptions, scope, analytical techniques, and  control strategies they can simulate or inform. Strictly speaking, the recent developments on mathematical model in this field have enhanced the capacity to predict the infection dynamics, evaluate efficacy of intervention, and inform public health policy. This review put together these advances, finds the persisting gaps in the mechanistic understanding and the data integration. Finally, it outlines few key directions for future studies aimed at refining   accuracy in prediction and  optimizing prevention. It is expected that a more comprehensive understanding of the interaction between pneumococcus and RSV would help develop more effective  ways to  prevent and manage the infection.

\subsection{Inclusion-Exclusion Criteria}
\textcolor{black}{The inclusion and exclusion criteria are defined to ensure that only relevant and high-quality studies are considered in this review. Since this review focuses on the advancements and developments in modeling both bacterial and viral pneumonia, we include studies that proposed, developed, or analyzed mathematical or computational models related to pneumonia dynamics. For both infections, we primarily focus on studies from the past fifteenth years related to disease dynamics and mathematical modeling. Although, we begin to explain the progression of modeling approach from very basic model such as SIR (Susceptible-Infected-Recovered) model. Four distinct databases, PubMed, Web of Science, Scopus, and Science Direct are used to search for scientific papers related to mathematical models developed for both bacterial and viral pneumonia. 
 In addition, studies focusing on co-infection involving these diseases are excluded from this review. Within-host models are also not considered in this study.}

 \subsection{Organization}
 The remainder of this review is structured as follows. In Section \ref{available vaccine}, we provide a detailed discussion on the development and the availability of vaccines for both bacterial and viral pneumonia. We also highlight their effectiveness, coverage rate, and implementation challenges in public. Section \ref{model} describes  the  epidemic modeling background, beginning with the classical SIR model and its extensions. The extensions  also include the stochastic modeling framework that incorporate the uncertainty and the real-world variability. This section also consider the vaccination epidemic models and their integration into the disease transmission dynamics. Finally, in Section \ref{conclusion}, we  give few key conclusions and outline some future research directions, focusing on the role of advanced modeling in public health interventions and policy-making.

\section{Vaccines Availability} \label{available vaccine}
Pneumococcal conjugate vaccines (PCVs) have been developed to provide protection against pneumococcal diseases caused by over 100 serotypes of bacteria \textit{S. pneumoniae}. A brief description of the evolution of PCV and its dosing
 strategies in England is available in the article of Choi \textit{et al.} \cite{2019_choi_estimated}. Similarly, RSV vaccines and monoclonal antibodies have been developed to provide protection against severe RSV disease, particularly in infants, young children, older adults, and other high-risk groups. A brief summary of the progress in RSV vaccine development and current immunization strategies can be found in the recent literature \cite{deng2025progress}.
\subsection{Vaccines Availability for Bacterial Infections Caused by \textit{S. Pneumoniae}}
Two types of pneumococcal vaccines are currently available: (i) pneumococcal polysaccharide vaccines (PPSV)  and (ii) pneumococcal
 conjugate vaccines. These vaccine  are designed to protect against infections caused by \textit{S. pneumoniae}, a bacterium responsible for pneumonia, meningitis, bacteremia, sinusitis, otitis media  and other serious illnesses. These vaccines target specific serotypes of the bacteria, reducing the risk of severe disease, especially in young children, the elderly, and immunocompromised individuals. Several PCVs are available, each covering different numbers of pneumococcal serotypes. However, since these vaccines target specific serotypes, there is a risk of replacement by non-vaccine serotypes over time. Current research focuses on developing serotype-independent vaccines, but this remains a complex and challenging task.
\subsubsection{PCVs and PPSV} 
These vaccines use a conjugated protein to improve immune response, especially in young children. Below, we present a summary of the currently available vaccines and their key characteristics.
\begin{enumerate}[(i)]
 \item \textbf{PCV7 (Prevnar 7):} It is a pneumococcal conjugate vaccine that covers seven serotypes (4, 6B, 9V, 14, 18C, 19F, and 23F); it was approved by the FDA in 2000 and has since been replaced by PCV13 \cite{2016_daniels,lee2014efficacy}.
    
    \item \textbf{PCV10 (Synflorix):}  This vaccine covers ten serotypes, including those in PCV7 plus serotypes 1, 5, and 7F, and was approved in December 2008 in Canada and in March 2009 by the European Medicines Agency \cite{2011_choi}.
    
    \item \textbf{PCV13 (Prevnar 13):} It provides protection against thirteen serotypes, comprising the PCV10 serotypes along with serotypes 3, 6A, and 19A, and was approved by the FDA in 2010 \cite{2016_daniels}.
    
    \item \textbf{PCV15 (Vaxneuvance):} This covers fifteen serotypes by including all PCV13 serotypes plus 22F and 33F, and was approved for use in adults on July 16, 2021 \cite{prasad2023public}.
    
    \item \textbf{PCV20 (Prevnar 20):} PCV20 targets twenty serotypes, consisting of all PCV15 serotypes along with 8, 10A, 11A, 12F, and 15B, and was approved in 2021 in the United States and in 2022 in the European Union \cite{2023_shirley}.
    
    \item \textbf{PPSV23 (Pneumovax 23):} This is a pneumococcal polysaccharide vaccine that covers twenty-three serotypes, including those in PCV20 and additional serotypes important for adult protection, and was approved by the FDA in 1983 \cite{2016_daniels}.
\end{enumerate}
\subsection{Vaccines Availability for Viral Infections Caused by RSV} 
According  to the Centers for Disease Control and Prevention \cite{CDC_RSV}, the following vaccines and monoclonal antibodies for RSV infection are available:
\begin{enumerate}[(i)]
    \item \textbf{Arexvy (GSK):} It is an adjuvanted vaccine based on the RSV prefusion F protein. Arexvy demonstrated approximately 77\% effectiveness against RSV-related emergency visits and 83\% effectiveness against hospitalizations in adults aged 60 and older during the 2023–2024 RSV season. Clinical trial data suggest that the  protection may persist for up to 23 months, although some waning  has been observed over time \cite{closegrading}.

\item \textbf{Abrysvo (Pfizer):} Abrysvo is a bivalent prefusion F protein vaccine covering both RSV-A and RSV-B strains. It shows around 79\% effectiveness in preventing emergency department encounters and 73\% effectiveness against hospitalizations for RSV in adults 60 and above. Clinical trials indicate durable immunity for approximately 18 months, with gradual decline in protection \cite{lassen2025bivalent}.

\item \textbf{mRESVIA (Moderna):} It is an mRNA-based RSV vaccine. Although real-world effectiveness data for mRESVIA are not yet available, clinical trials showed ~80\% efficacy against symptomatic RSV infection during the first 4 months. Efficacy declined to ~56\% over the first year following a single dose \cite{wilson2023efficacy}.

\item \textbf{Monoclonal antibodies:}
	For infants, RSV prevention primarily involves long-acting monoclonal antibodies (mAbs), which are not vaccines. These products provide passive immunity for several months without inducing immune memory. Nirsevimab (Beyfortus) and Clesrovimab (Enflonsia) are considered as available monoclonal antibodies. These monoclonal antibodies are administered at a particular age as intramuscular injections and typically provide protection for approximately five months, covering one RSV season \cite{ananworanich2021bringing}.

\item \textbf{Maternal vaccination for newborn protection:}
	Maternal vaccination with Pfizer’s Abrysvo administered as a single dose between 32 and 36 weeks of gestation allows maternal antibodies to cross the placenta and provide protection to the infant during the first months of life \cite{dhanvi2024use}.
\end{enumerate}

\color{black}
\section{Modeling Background} \label{model}
Various mathematical models are used to study pneumonia caused by \textit{S. pneumoniae} and RSV. Beyond   infection-dynamics model based on ordinary differential equations (ODEs) and stochastic differential equation (SDEs), a variety of other modeling frameworks have been developed in the literature. These include partial differential equation (PDE) models to describe spatial spread in the lungs, delay models to capture incubation or immune-response delays, agent-based models to investigate cell–pathogen interactions, etc. However, in this work, we mainly focus on ODE-based deterministic models and stochastic models.
\subsection{Deterministic Epidemic Models}
An epidemic model  mathematically describes the spread of infectious diseases in a population. One of the basic and simplest mathematical modeling for pneumonia disease is Susceptible-Infected-Recovered (SIR) framework. The SIR model, proposed by Kermack and McKendrick \cite{1997_Kermack},  is the most widely used  model for analyzing epidemics at  population level where the compartments can be limited within thrice only and they are (i) susceptible $(S(t))$, (ii) infected $(I(t))$, and  (iii) recovered $(R(t))$ individuals. The mathematical formulation of the model is given by the following system of ODEs:
\begin{equation} \label{model_SIR}
\left.
    \begin{aligned}
\frac{dS(t)}{dt} &= -\beta S(t) I(t), \\
\frac{dI(t)}{dt} &= \beta S(t) I(t) - \gamma I(t), \\
\frac{dR(t)}{dt} &= \gamma I(t),
\end{aligned}
\right\}
\end{equation}
subject to non-negative initial conditions $S(0)\geq 0, I(0)\geq 0, R(0)\geq 0$. Here, 
\begin{itemize}
    \item \( \beta \) denotes the \textbf{disease transmission rate} (rate at which susceptibles become infected), and
    \item \( \gamma \) indicates  the \textbf{recovery rate} (rate at which infected individuals recover and gain immunity) \cite{1997_Kermack}.
\end{itemize}
In model~\eqref{model_SIR}, the parameters $\beta$ and $\gamma$ are generally assumed to be constant. 
The simple SIR model effectively captures basic epidemic thresholds and peak dynamics through its analytical power. Many extensions of this simple SIR model \eqref{model_SIR} have been developed to account for the  epidemiological complexities of particular pneumonia infection, such as latency periods, loss or waning of immunity, age structure, spatial and spatio-temporal heterogeneity, seasonality, vaccination \cite{2008_Temime_1}, demographic changes, the roles of carriers \cite{2013_Otieno}, natural immunity \cite{2024_purohit_novel}, and control strategies \cite{2021_swai_optimal,2017_Tilahun} to better capture realistic disease transmission dynamics. 
Since pneumonia is highly seasonal, with peaks occurring in winter, a common and important extension involves incorporating \textit{seasonal forcing} into the transmission rate $(\beta\equiv\beta(t))$ that can be expressed in the following form
\begin{align} \label{betat}
    \beta(t) = b_0  + b_1 \cos(2\pi t + \phi),
\end{align}
where the parameter $b_0$ is the average transmission rate, $b_1$ modulates the amplitude of seasonal variation, and $\phi$ controls the phase shift \cite{2001_weber_modeling}. Several other similar forms of seasonality parameters are available in \cite{li2022analysis, el2023periodic, yang2016seasonality}. This approach is particularly relevant for respiratory diseases that mostly driven by winter seasonality, such as bacterial pneumonia, viral pneumonia, etc. 
Weber \textit{et al.}~\cite{2001_weber_modeling} applied this seasonal variability (shown in equation \eqref{betat}) in the parameter disease transmission rate \(\beta(t)\) to a   Susceptible-Infected-Recovered-Susceptible (SIRS) framework to   see  the patterns in the RSV epidemic. Thay included both annual and biennial cycles across different regions. It is important to note that this model estimated the basic reproduction number $(\mathcal{R}_0$, to range from 1.2 to 2.1. In order to better capture  the  complex re-infection dynamics of RSV, the authors also developed an MSEIRS4 model, introducing additional compartments for maternal immunity ($M$), latent infection ($E$), and four reinfection susceptibility classes ($S1, S2, S3, S4$), each with reduced transmission rates \(\beta_2 = 0.5\beta_1\), \(\beta_3 = 0.35\beta_1\), \(\beta_4 = 0.25\beta_1\) where  $\beta_1$, $\beta_2$, $\beta_3$ and $\beta_4$ represent the transmission rates for individuals experiencing their first, second, third and and fourth  infections, respectively, reflecting the decreasing susceptibility to RSV following repeated exposures.
by fitting this model to a hospitalization data through the non-linear least squares method resulted in higher value of \(\mathcal{R}_0\), lies between 5.4 and 7.1. Note that this variability in the estimated  value of $\mathcal{R}_0$ highlights the sensitivity of \(\mathcal{R}_0\) to model structure, assumption and, in particular, the inclusion of reinfection pathways. Both models highlight seasonality as a key driver of epidemic breaks, though inconsistent associations with climate factors suggest the need for region-specific environmental analyses and further structural validation.
The seasonal epidemics of RSV in young children were also studied using a two-age structured SEIR model and fitting to a population-level RSV dataset in \cite{moore2014modelling}. The authors of the study \cite{moore2014modelling} found a distinct biennial seasonal pattern, characterized by alternating peak sizes during the winter months.

In order to develop cost-effective strategies for pneumonia control, Tilahun \textit{et al.} \cite{2017_Tilahun}  modified the SIR model \eqref{model_SIR} and proposed another  dynamic model (equation \eqref{model_Tilahun}) in 2017.  The proposed model is given by the following system of nonlinear coupled ODEs with minor changes in notations:
\begin{equation}\label{model_Tilahun}
\left.
\begin{aligned}
    \frac{dS(t)}{dt} &= (1 - p)\pi + \phi V(t) + \delta R(t) - (\mu + \lambda + \vartheta) S(t), \\
     \frac{dI(t)}{dt} &= (1 - \rho) \lambda S(t) + (1 - \rho) \epsilon \lambda V(t) + \chi C(t) - (\mu + \alpha + \eta) I(t), \\
    \frac{dC(t)}{dt} &= \rho \lambda S(t) + \rho \epsilon \lambda V(t) + (1 - q) \eta I(t) - (\mu + \xi + \chi) C(t), \\
    \frac{dR(t)}{dt} &= \xi C(t) + q \eta I(t) - (\mu + \delta) R(t),\\
     \frac{dV(t)}{dt} &= p\pi + \vartheta S(t) - (\mu + \epsilon \lambda + \phi) V(t)
\end{aligned}
\right\}
\end{equation}
subject to non-negative initial conditions. 
Here,  $V(t)$ and $C(t)$ are two newly introduced compartments of the model \eqref{model_Tilahun} and represents the vaccinated and carrier classes, respectively. This model assumes that a fraction $p$ of the population is vaccinated before the disease outbreak, while the remaining fraction $(1-p)$ is susceptible, and the susceptible class is increased by vaccine non-responders from the vaccinated class $V(t)$ at the waning rate $\phi$ and by individuals losing temporary immunity from the recovered class $R(t)$ at rate $\delta$. Tilahun \textit{et al.} \cite{2017_Tilahun} found through numerical experiments that combining prevention and treatment yields the most cost-effective outcome, balancing both cost and health benefits. They used the Incremental Cost-Effectiveness Ratio (ICER), defined as
$$\text{ICER} =\dfrac{\text{Difference in costs between strategies}}{\text{Difference in health effects between strategies}}.$$
to compare the relative efficiency of different intervention strategies.
Building upon previous work, White \textit{et al.} \cite{2005_white_transmission} developed a 12-compartment model to explore the transmission dynamics of RSV groups A and B (classified based on the reactions with panels of monoclonal
 antibodies) in England \& Wales, Turku, and  Finland. The model incorporates a range of epidemiologically relevant characteristics, such as seasonal forcing, primary and secondary infections, homologous and heterologous immunity, and antigenic variability in  G glycoprotein. The force of infection for group $i$ is defined as
$$
\lambda_i = \beta_i(t)(P_i + \eta(P_{ji} + Y_i + Y_{ji})), (i,j) \in \{\text{(A,B), (B,A)}\}
$$
with $\beta_A(t) = b_A(1 + a \cos(2\pi(t - \phi)))$ and $\beta_B(t) = r\beta_A(t)$, where $r$ reflects group $B$’s relative transmissibility. In their model, $P_i,\; i \in \{\mathrm{A,B}\},$ denotes the proportion of hosts infected with group $i$ without prior infection. 
$P_{BA}$ ($P_{AB}$) denotes infection with group A (B) following a previous infection with group B (A). 
$Y_A$ and $Y_B$ denote infection with the same group following a previous infection with group A and B, respectively, 
while $Y_{BA}$ and $Y_{AB}$ denote infection with group A and B, respectively, following previous infections with both groups. The parameters $\beta_A$ and $\beta_B$ denote the seasonally varying transmission coefficients for groups A and B, respectively, where $b_A$ denotes the average transmission coefficient for group A, $a$ denotes the amplitude of the transmission coefficient and and $\phi$ represents the phase shift. The parameters of the model were estimated via the simplex method to fit hospitalization data. The model captured 6-year epidemic cycles in England \& Wales and 4-year cycles in Finland. Estimated $\mathcal{R}_0$ for group A is 2.81 and 2.46 in England \& Wales and Finland, respectively. The study illustrates how  transmissibility differences and partial cross-protection influence strain coexistence and epidemic patterns.

In this direction, Acedo \textit{et al.} \cite{acedo2010} proposed an age-structured SIRS model to evaluate neonatal RSV vaccination strategies in Valencia, Spain, stratifying the population into infants ($<$1 year) and older individuals. In their study,  the seasonal transmission is represented by the time-dependent parameter $\beta(t)$, given in equation \eqref{betat}.
The model parameters are calibrated to 2001–2004 weekly hospitalization data via the \textit{Nelder–Mead} algorithm. At 85\% vaccine coverage, infant infections were reduced by approximately 63\% within two years. A cost-effectiveness analysis, incorporating vaccination, hospitalization, and productivity costs, projected net savings of €2 million.

On the other hand,  considering of one sequence type associated with two serotypes Lamb \textit{et al.} \cite{2011_lamb_simple} proposed an epidemic model for this bacterial pneumonia caused by \textit{S. Pneumoniae} and the following system of ODEs with slight symbolic modification describes  it: 
\begin{equation}\label{model_lamb}
\left.
\begin{aligned}
    \frac{dX(t)}{dt} &= (1 - p)\pi - uX(t) - \beta_1 X(t)(T_1(t) + V_{T_1}(t)) + \gamma T_1(t), \\
    \frac{dT_1(t)}{dt} &= \beta_1 X(t)(T_1(t) + V_{T_1}(t)) - (\gamma + u)T_1(t), \\
    \frac{dV_1(t)}{dt} &= p\pi - uV_1(t) - \beta_1 V_1(t)(T_1(t) + V_{T_1}(t)) + \gamma V_{T_1}(t), \\
    \frac{dV_{T_1}(t)}{dt} &= \beta_1 V_1(t)(T_1(t) + V_{T_1}(t)) - (\gamma + u)V_{T_1}(t)
\end{aligned}
\right\}
\end{equation} subject to non-negative initial conditions. 
In this model \eqref{model_lamb}, the total  population is categorized into four groups: those who are unvaccinated and susceptible to carriage of sequence type 1 $(X(t))$, those who are unvaccinated but carrying sequence type 1 $(T_1(t))$, those who are vaccinated and susceptible to carriage of sequence type 1 $(V_1(t))$, and those who are vaccinated while carrying sequence type 1 $(V_{T_1}(t))$. This modeling framework is slightly different from the fundamental SIR model given in equation  \eqref{model_SIR}. It is important to note that the susceptible compartment $X(t)$ in model~\eqref{model_lamb} is different from the susceptible compartment $S(t)$ in models~\eqref{model_SIR} and~\eqref{model_Tilahun}. In this model, the vaccine is assumed to fully protect against serotype 1 but provides no protection against serotype 2. Lamb \textit{et al.} \cite{2011_lamb_simple} explored  the relationship between sequence types and serotypes where a sequence type is able to manifest itself in one vaccine serotype and one non-vaccine serotype. While the effective reproduction number ($R_e$) for \textit{S. pneumoniae} was estimated to be approximately 1.49 in \cite{farrington2004reproduction}, whereas another study \cite{zhang2004mucosal} reported it to range between 1.8 and 2.2. However,  they proceeded with $R_e = 1.5$ for their analysis. However, although some model parameters such as $\pi$ and $p$ share the same notation in both models \eqref{model_Tilahun} and \eqref{model_lamb}, their numerical values are different.

Besides,  based on the assumptions (i) newborns are given additional dose of vaccine to elicit booster optimal levels of response, and (ii) all treated individuals get vaccinated after completing the dose, Kizito and Tumwiine \cite{2018_kizito_mathematical}  proposed a four-compartmental population dynamics model on pneumonia  given by the following system of ODEs:
\begin{equation}
\left.
\begin{aligned}
\frac{dS(t)}{dt} &= \Lambda - (\alpha + \mu) S(t) + \eta R(t), \\
\frac{dC(t)}{dt} &= \alpha \theta S(t) - (\mu + \xi + \pi) C(t), \\
\frac{dI(t)}{dt} &= \alpha (1 - \theta) S(t) + \pi C(t) - (\tau + \mu + \sigma) I(t), \\
\frac{dR(t)}{dt} &= \xi C(t) + \tau I(t) - (\mu + \eta) R(t)
\end{aligned}
\right\}
\end{equation} subject to non-negative initial conditions.  One of the key findings of their study is that pneumonia can be eradicated only when treatment and vaccination are implemented together, whereas treatment alone is insufficient, allowing the disease to persist in the population. Several other contributions in this direction are reported in \cite{2022_kotola_mathematical,ossaiugbo2021mathematical,almutairi2025optimal}.

In most epidemiological models, the demographic structure is incorporated mainly through age stratification. Typically, models use finer age divisions for young children ($<$5 years) who are most vulnerable and broader groupings for older ages. Only a few models (\textit{e.g.}, Campbell \textit{et al.} \cite{campbell2020modelling}; Kombe \textit{et al.} \cite{kombe2019model}; Poletti \textit{et al.} \cite{poletti2015evaluating}; Ramjith \textit{et al.} \cite{ramjith2021flexible},  Nerurkar et al \cite{nerurkar2025predicting}) use agent-based frameworks or focus specifically on children under two years (Hogan \textit{et al.} \cite{hogan2016exploring}; Paynter \cite{paynter2016incorporating}). Transitions between age groups are usually modeled at rates proportional to the inverse of the age-band width, though some models apply more complex aging mechanisms to preserve realistic population structures. In addition to age categories, a few models also include, shool, household, and geographical stratification to capture finer demographic or  heterogeneity in the contact-pattern \cite{poletti2015evaluating}. Therefore, epidemic modeling of pneumonia is useful  to understand its spread and implement the effective control strategies. It is worth nothing that the accurate models  always support timely interventions, by reducing morbidity and mortality rates.

\subsection{Stochastic Epidemic Models}
Stochastic models provide a principled framework to address the discrepancy between observed data and the underlying epidemic process.
Instead of empirically adjustments to the observed data for deterministic models, a statistical approach offer a better way to account the uncertainty in both disease transmission and observation processes. In this context, the state-space models \cite{durbin2012} provides a flexible framework for jointly modeling unobserved epidemic states and noisy observations.
In general, the state-space models have  two components:
\begin{itemize}
    \item \textbf{State model}: It describes the temporal evolution of unobserved epidemic states, $X_t \sim p(X_t \mid X_{t-1}, \theta)$.
    \item \textbf{Observation model}: It links these states to observed data, $Y_t \sim p(Y_t \mid X_t, \theta)$,
\end{itemize}
where $X_t$ represents the latent state (\textit{e.g.}, true infections), $Y_t$ denotes the observations (\textit{e.g.}, hospitalizations), and $\theta$ is the parameters. 
This framework accommodates various epidemiological scenarios. For instance, the state models can represent SIR dynamics with stochastic transitions, while the observation models can account for hospital admissions as a function of new infections, incorporating time lags and under-reporting. This approach captures both epidemic dynamics and data collection imperfections.

It is observed that a primary objective in epidemiological modeling is estimating quantities like the basic reproduction number $(R_0)$ or effective reproduction number $(R_t)$ (time-varying), often nonlinear functions of model parameters (\textit{e.g.}, $R_0 = \dfrac{\beta}{\gamma}$ for SIR model). Inferring $R_t$ from partially observed data is well-suited to the state-space framework. Moreover, components like the transmission function $\beta(t)$ can be modeled flexibly using nonparametric or semi-parametric approaches \cite{bouman2024, zhou2020, andrade2022}.
Thus, state-space models bridge mechanistic disease spread models with the stochasticity of real-world data, enabling robust inference.

Moreover, the Bayesian framework naturally extends this nonparametric or semi-parametric  approach by coherently representing and propagating uncertainties in parameters, latent states, and observation processes. It provides posterior distributions for quantities of interest, facilitating probabilistic forecasting and decision-making. Additionally, it supports prior knowledge incorporation, model comparison, and hierarchical modeling, which are critical in data-sparse or noisy settings. 
This leads to the Bayesian workflow \cite{gelman2020, gelman2006, gelman2013}, encompassing model building, prior specification, computation, model checking, and sensitivity analysis, essential for reliable inference in complex epidemiological contexts. Strictly speaking, the Bayesian inference workflow \cite{gelman2020, gelman2013, gelman2006} is an iterative process for building, fitting, and critiquing probabilistic models. It provides a principled framework for incorporating domain knowledge through prior distributions, estimating parameters and latent states from noisy observations, and quantifying uncertainty in a coherent manner. 

In this spirit, let us now begin with the Bayesian discrete-time stochastic SIRS model proposed by \cite{corberan2014} to analyze RSV dynamics among children younger than two years of age based on weekly hospitalization data in Valencia, Spain from 2001 to 2004. The model accurately captured disease patterns and predicted the onset and progression of new RSV epidemics. For defining observation model, let \( y_t \) be the number of RSV hospitalizations among children under 2 years of age reported at week \( t \) (\( t = 1, 2, \dots, T \)). At the first level of the model, it is assumed that \( y_t \) follows a Binomial distribution: $y_t \sim \text{Binomial} (i_t, \rho)$
where \( i_t \) is the number of new infections at time \( t \) and \( \rho \) is the probability of hospitalization given infection which is assumed to be 1 for the entire analysis due to non-availability of a statistical robust estimate. The number of new infections \( i_t \) is modeled as $i_t \sim \text{Binomial} (S_{t-1}, p_t)$, where \( S_{t-1} \) represents the susceptible population at time \( (t-1) \), and \( p_t \) is the probability of becoming infected at time \( t \). Although other discrete distributions like Poisson distribution can also be used for modeling counts, the Binomial distribution is preferred here due to the finite size of the children population.

For the process model (see \cite{corberan2014} for details), the model conceptualizes the population as susceptible ($S$), infected ($I$), and recovered ($R$) compartments with transitions modeled as a discrete-time Markov chain. In this discrete-time model, the number of individuals in each compartment is examined at discrete time steps. Using a fixed population, the number of susceptible, infected, and recovered individuals at time $t$ are updated through the following recursive equations:

\begin{equation}
\left.
\begin{aligned}
S_t &= S_{t-1} - i_t + bR_{t-1} + n_t - \frac{S_{t-1}}{N} n_t, \\
I_t &= I_{t-1} - aI_{t-1} + i_t - \frac{I_{t-1}}{N} n_t, \\
R_t &= R_{t-1} - bR_{t-1} + aI_{t-1} - \frac{R_{t-1}}{N} n_t
\end{aligned}
\right\}
\end{equation} subject to non-negative initial conditions, where $a$ is the proportion of infected individuals that recover per unit time; $b$ is the proportion of recovered individuals who lose their immunity and become susceptible again per unit time; $n_t$ is the number of births at time $t$; and $N$ is the constant population size.
Given that the average recovery time from RSV is 10 days and immunity lasts around 200 days, the recovery rate is set at $a=0.7$ (per week) and the immunity loss rate at $b=0.035$ (per week), treating these as fixed parameters based on established estimates \cite{acedo2010}.
Additionally, $n_t$ and $N$  are obtained using demographic data from Spain's National Institute of Statistics. The infection probability $p_t$ is modeled as:
$$
p_t = \min \left\{ \frac{i_{t-1}^{\alpha} \exp\{r_t\}}{1 + i_{t-1}^{\alpha} \exp\{r_t\}} + c, \; 1 \right\},
$$
where \( i_{t-1}^\alpha \) captures the influence of the number of infectives at the previous time step with heterogeneous mixing (\( \alpha = 1 \) implies mass action, \( \alpha < 1 \) indicates sublinear dependence), \( \exp\{r_t\} \) represents the time-varying transmission rate, and \( c \) is a small constant accounting for external infections that enables the model to explain epidemic resurgence after fade-out ($i_{t-1}=0$), which is a purely infective-dependent model. The function \(\min\{\cdot, 1\}\) ensures \( p_t \) remains within \([0, 1]\) to maintain a valid probability. This form arises from modeling the logit of \( p_t \) as
\[
\log\left( \frac{p_t}{1-p_t} \right) = \alpha \log(i_{t-1}) + r_t.
\]
In order to model seasonal variation in transmission, $r_t$ is expressed using a sine-cosine series:
\[
r_t = a_0 + \sum_{k=1}^{K} \left[ a_{2k-1} \sin\left( \frac{2k\pi t}{52} \right) + a_{2k} \cos\left( \frac{2k\pi t}{52} \right) \right] + \epsilon_t
\]
where $\epsilon_t$ captures unexplained weekly variability. The choice of $K$ is data-driven, selected as the largest $k^*$ such that $a_{2k^*-1}$ and $a_{2k^*}$ are significant. Adding random effects in the transmission rate model accounts for over-dispersion providing a more realistic description of the transmission pattern.

In \cite{corberan2014}, to demonstrate performance, four alternative discrete-time SIRS model formulations (two stochastic with seasonal and constant transmission rates and their corresponding deterministic versions) were studied. In the course of numerical studies, they illustrate the model's fit to weekly RSV hospitalization data from July 2, 2001, to June 21, 2004. The posterior means and 95\% confidence intervals closely align with the observed data, indicating a precise representation of the disease pattern over this period. In addition to this, fitted model is used to obtain forecasts for the last 26 weeks of 2004 (week 27 to week 52). The model accurately predicts the timing and magnitude of the 2004 RSV epidemic, achieving a lower forecast Root Mean Squared Error (RMSE) than alternative models, and effectively capturing the start and rise of the outbreak. Unlike deterministic models, which fail to predict epidemic onset when the number of infectives is zero (\textit{e.g.}, $y_T=0$), the stochastic nature of this model, combined with a constant infection probability $c$, enables it to anticipate new outbreaks, enhancing its predictive power.

In the numerical studies of the aforementioned work, the prior distributions are assigned to key parameters to encode prior knowledge and facilitate data-driven inference. A $\text{Uniform}~(0,1)$ prior is chosen for the mixing parameter ($\alpha$), reflecting uncertainty in population mixing from sublinear ($\alpha < 1$) to mass action ($\alpha = 1$). Posterior mean (0.85) suggests near-proportional mixing. The constant infection probability ($c$) is kept small and is assigned a $\text{Uniform}~(0,0.01)$ prior to ensure that it does not dominate the primary transmission term \( i_{t-1} \alpha \exp\{r_t\} \), reflecting minor external contributions. The posterior mean \( 1.46 \times 10^{-5} \) aligns with this intent. Next, transmission rate coefficients ($\{a_k\}$) are modeled using zero-mean Gaussian priors, with their standard deviations ($\sigma_{a_k}$) following $\text{Uniform}~(0, 5)$ distributions to allow flexible but stable seasonal variation in transmission. Similarly, random effects ($\{\epsilon_t\}$) are modeled with zero-mean Gaussian priors and Uniform priors on their standard deviations to capture unstructured variability without overfitting. These choices follow \cite{gelman2006} with the aim to providing weak information to stabilize variance estimates without overly restricting them. For details, kindly see Table \ref{tab:model-summary}. The article specifies $K=2$ to effectively represent RSV's annual seasonality without unnecessary complexity. With a 52-week year, $K=2$ includes terms with periods of 52 weeks (annual cycle) and 26 weeks (semi-annual cycle), capturing RSV's strong winter peaks. 

Later, Marc \textit{et al.} \cite{jornet2017} further extended this work to evaluate a vaccination strategy targeting newborns using the same formulation and Bayesian methodology. For the vaccination extension, they introduce an additional vaccinated  compartment ($V$) and parameter ($\nu$) representing the proportion of vaccinated newborns. The model shows significant reductions in hospitalizations with increasing vaccination coverage --- at 20\% coverage, hospitalization costs drop from €13.2M to €8.5M, and at 80\% coverage to just €2.5M. 

\begin{table}[ht]
\centering
\begin{tabular}{|p{1.0cm}|p{3.3cm}|p{2.8cm}|p{4.5cm}|p{1.2cm}|}
\hline
\textbf{Source} & \textbf{Observation Model} & \textbf{Process } & \(\boldsymbol{\beta}\) \textbf{(Transmission Rate)} & \textbf{Disease} \\
\hline
\cite{andrade2020} & Binomial & Age-stratified SEIR & Constant & Influenza \\
\hline
\cite{andrade2021} & Poisson & $SEIR$ & Constant & Influenza \\
\hline
\cite{andrade2022} & Poisson, Negative Binomial & $SEI3R$ & Functions with non-parametric priors - Geometric Brownian motion, Cox-Ingersoll-Ross and exponential smoothing & Covid-19 \\
\hline
\cite{andrade2023} & Negative Binomial & $SE^iI^jR$ & Constant & Influenza \\
\hline
\cite{osthus2017} & Beta & Dirichlet distribution with SIR dynamics & Constant & Influenza \\
\hline
\cite{grinsztajn2021} & Negative Binomial & $SIR/SEIR$ & Constant/time-varying with   & Influenza \\
\hline
\end{tabular}
\caption{Summary of observation and process models with transmission rate estimates.}
\label{tab:model-summary}
\end{table}

{\color{black} A few years ago, \cite{ARENAS2009206} studied the dynamics of RSV transmission in the population using stochastic models. They developed stochastic perturbations on the demographic parameter in the stochastic model as well as on the transmission rate of the RSV. Strictly speaking, they considered two models using stochastic perturbations on some parameters as in \cite{SAHA2008458}. The first one studies perturbations on the birth rate, and the second one is based on the baseline transmission parameter. The both cases in general can be represented as the Itô type stochastic differential system :  
$$dX(t) = f (t, X(t)) dt + g(t, X(t)) dW(t)$$ with the initial condition $X(t_0)=X_0, t\in[t_0,t_f]$, where $X(t)=(S(t),I(t),R(t))^T$, and the solution $\{X(t), t\in[t_0,t_f]\}$ is an Itô process. Here $f$ is a non-stochastic continuous function, and $g$ is also a continuous function of random element $X(t)$ and time $t$. Moreover, $W(t)$ is a certain Wiener process.  

Now, in the first case, suppose that $\mu$ denotes the birth rate associated with RSV spread dynamics. Observe that RSV illness depends on the timing and severity of outbreaks in a community vary from year to year, and hence, the birth rate is also varying in a varying population in every month/year. In this case, \cite{ARENAS2009206} investigated  the effect of birth rate perturbations on the dynamics of the RSV model as even a small change can produce significant dynamical changes. Let $\tilde{\mu}$ is the perturbed birth rate, and $$\tilde{\mu} (t) = \mu(t) + \alpha W(t),$$ where $\alpha\in\mathbb{R}$ is an arbitrary constant, and $W(t)$ is a standard Wiener process. Non technically speaking, the birth parameter fluctuates randomly around some constant average value $\mu$ assumed in the non-stochastic model, and these fluctuations may be due to several factors such weather, economic conditions and others. Consequently, using the relation between $(\tilde\mu)$ and $\mu$, the stochastic differential system model with perturbed birth rate is of form:
\begin{equation}
\left.
\begin{aligned}
dS(t) &= [(\mu - \mu S(t) - \beta(t)S(t)I(t) + \gamma R(t))] dt +  \alpha (1 - S(t)) dW(t),\\
dI(t) &= [(\beta(t)S(t)I(t) - \nu I(t) - \mu I(t))] dt - \alpha I(t) dW(t),\\
dR(t) &= [\nu I(t) - \mu R(t) - \gamma R(t)]dt - \alpha R(t) dW(t)
\end{aligned}
\right\}
\end{equation} with non-negative initial conditions.
In the next case, due to the same reason described in the preceding paragraph, the baseline transmission parameter, which is denoted by $b_{0}$, given in equation \eqref{betat}, may have stochastic/random perturbation. Let $\tilde{b}_{0}$ be the perturbed baseline transmission parameter, and $$\tilde{b}_{0} (t) = b_{0} (t) + \alpha W(t),$$ where $\alpha$ and $W(t)$ are the same as defined in the preceding paragraph. Here also, non-technically speaking, the baseline transmission parameter $b_{0}$ is equal to a constant average value in addition to a time fluctuating term, and these term follows a normal distribution with mean zero. Therefore, the stochastic differential system model with perturbed transmission parameter is of the form: 
\begin{equation}
\left.
\begin{aligned}
dS(t) &= [(\mu - \mu S(t) - \beta(t)S(t)I(t) + \gamma R(t))] dt - \frac{\alpha\beta(t)}{b_{0}}S(t) I(t) dW(t),\\
dI(t) &= [(\beta(t)S(t)I(t) - \nu I(t) - \mu I(t))] dt + \frac{\alpha\beta(t)}{b_{0}}S(t) I(t) dW(t),\\
dR(t) &= [\nu I(t) - \mu R(t) - \gamma R(t)]dt
\end{aligned}
\right\}
\end{equation}} with non-negative initial conditions. 

Finally, in the context of disease modeling, it is an appropriate place to mention that recently \cite{Ghosh2025} studied estimation of time-varying recovery and death rates from epidemiological data using the similar approach of the Nararaya-Watson estimator (see \cite{MR172400} and \cite{MR185765}) in estimating the unknown non-parametric regression function. In the course of this study, the authors also implemented the proposed methodology on the data set associated with a few VPDs. Besides, \cite{MR4730405} 
investigated modeling of parasite dynamics using a stochastic simulation-based approach and parameter estimation via modified sequential-type approximate Bayesian computation, and \cite{MR4728823} studied the memory effects in disease modeling through kernel estimates with oscillatory time history. A few years ago, \cite{MR3882131} proposed a geostatistical methods for disease mapping and visualization using data from spatio-temporally referenced prevalence surveys, and \cite{MR3687968} developed network models of infectious disease spread. During almost the same time, \cite{MR3693541} developed and applied a novel stochastic compartmental model to a large dataset on Clostridium Difficile Infection (CDI) in three Oxfordshire hospitals over a 2.5 year period which combines genetic information on 858 confirmed cases of CDI with a database of 750,000 patient records, and for communicable diseases, \cite{MR3480501} reconstructed transmission trees using densely sampled genetic data. Recently, \cite{MR4176186} reviewed various multi-compartment infectious disease models, but it is entirely different from the present work in various senses. 
All of these are relevant statistical methodologies associated with disease modeling.

\subsection{Vaccination Models}
\begin{figure}[h!]
    \centering
    \includegraphics[width=0.99\linewidth]{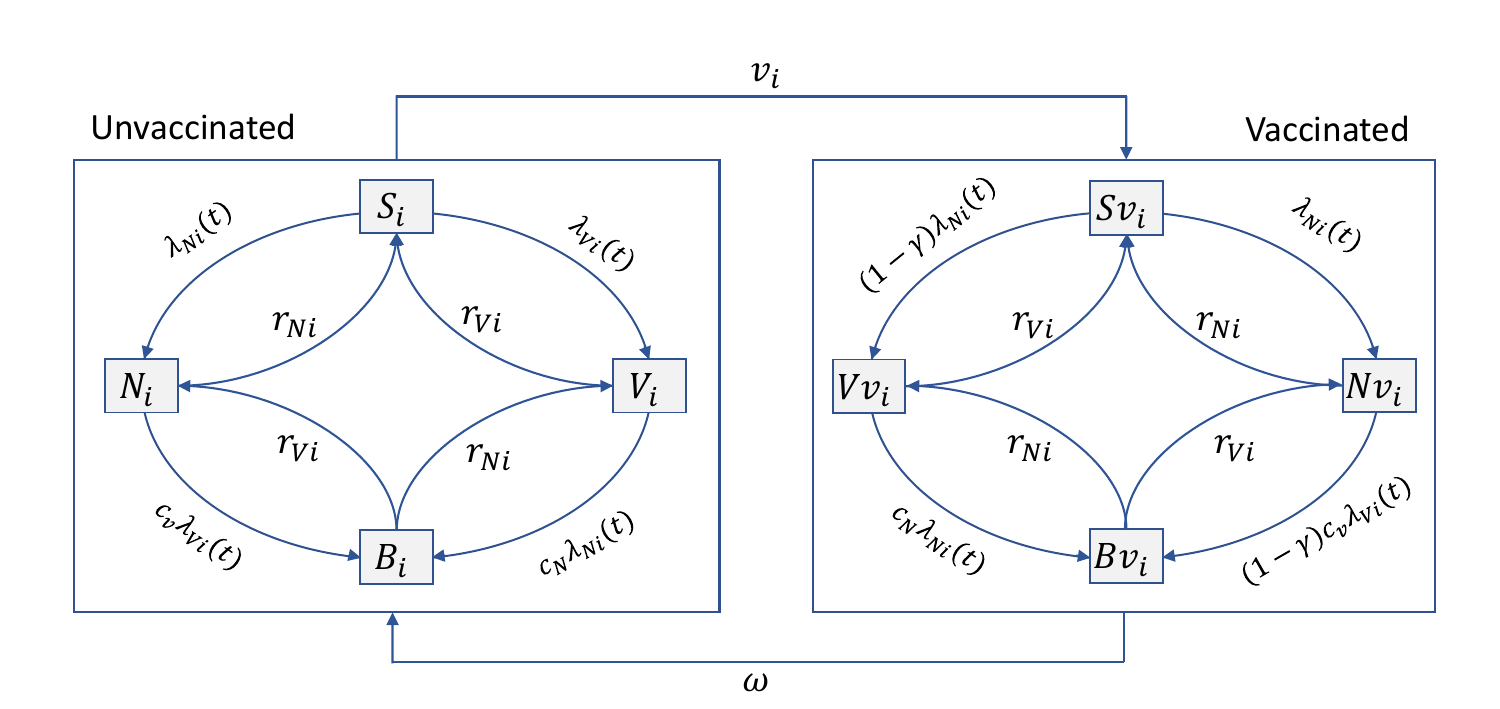}
    \caption{$S:$ unvaccinated susceptible; $S_v:$ susceptible vaccinated; $V:$ unvaccinated carriers of pneumococcal serotypes contained in PCV7 or serotype 6A (vaccine serotype group, VT); $Vv:$ vaccinated carriers of pneumococcal serotypes contained in PCV7 or serotype 6A; $N:$ unvaccinated carriers of pneumococcal serotypes not contained in PCV7 (non-vaccine serotype group, NVT); $Nv:$ vaccinated carriers of pneumococcal serotypes not contained in PCV7; $i$: age; t: time.}
    \label{fig:melegaro}
\end{figure}

Epidemic vaccine models are essential for understanding the impacts of vaccines on public health. These models help predict disease transmission, assess vaccine effectiveness, and evaluate changes in disease patterns over time, etc. These models also analyze the factors influencing vaccine coverage and potential shifts in viral and bacterial strains. Different kind of mathematical models incorporating the impacts of vaccines are available in the existing literature \cite{ 2025_ALDILA_100394,2017_Tilahun,2012_greenhalgh_mathematical,2021_swai_optimal}. Using epidemiological data from England and Wales and pre and post-vaccination surveillance data from the USA, Melegaro \textit{et al.} \cite{2010_melegaro_dynamic}  developed the following age-structured, compartmental model:

\begin{equation} \label{vaccine_model}
\left.
    \begin{aligned}
\frac{dS_i(t)}{dt} &= r_{Vi}\,V_i(t) + r_{Ni}\,N_i(t)
 - S_i(t)\bigl(\lambda_{Vi}(t) + \lambda_{Ni}(t)\bigr)
 - \pi_i(t)\,S_i(t) + \omega\,Sv_{i}(t), \\[4pt]
\frac{dV_i(t)}{dt} &= S_i(t)\,\lambda_{Vi}(t)
 - c_N\,\lambda_{Ni}(t)\,V_i(t)
 - r_{Vi}\,V_i(t) + r_{Ni}\,B_i(t)
 - \pi_i(t)\,V_i(t) + \omega\,Vv_{i}(t), \\[4pt]
\frac{dN_i(t)}{dt} &= S_i(t)\,\lambda_{Ni}(t)
 - c_V\,\lambda_{Vi}(t)\,N_i(t)
 - r_{Ni}\,N_i(t) + r_{Vi}\,B_i(t)
 - \pi_i(t)\,N_i(t) + \omega\,Nv_{i}(t), \\[4pt]
\frac{dB_i(t)}{dt} &= c_N\,\lambda_{Ni}(t)\,V_i(t)
 + c_V\,\lambda_{Vi}(t)\,N_i(t)
 - B_i(t)\,(r_{Ni}+r_{Vi})
 - \pi_i(t)\,B_i(t) + \omega\,Bv_{i}(t), \\[8pt]
\frac{dSv_{i}(t)}{dt} &= r_{Vi}\,Vv_{i}(t) + r_{Ni}\,Nv_{i}(t)
 - Sv_{i}(t)\bigl((1-\gamma)\,\lambda_{Vi}(t) + \lambda_{Ni}(t)\bigr)
 + \pi_i(t)\,S_i(t) - \omega\,Sv_{i}(t), \\[4pt]
\frac{dVv_{i}(t)}{dt} &= Sv_{i}(t)\,(1-\gamma)\,\lambda_{Vi}(t)
 - c_N\,\lambda_{Ni}(t)\,Vv_{i}(t)
 - r_{Vi}\,Vv_{i}(t) + r_{Ni}\,Bv_{i}(t)
 + \pi_i(t)\,V_i(t) - \omega\,Vv_{i}(t), \\[4pt]
\frac{dNv_{i}(t)}{dt} &= Sv_{i}(t)\,\lambda_{Ni}(t)
 - c_V\,(1-\gamma)\,\lambda_{Vi}(t)\,Nv_{i}(t)
 - r_{Ni}\,Nv_{i}(t) + r_{Vi}\,Bv_{i}(t)
 + \pi_i(t)\,N_i(t) - \omega\,Nv_{i}(t), \\[4pt]
\frac{dBv_{i}(t)}{dt} &= c_N\,\lambda_{Ni}(t)\,Vv_{i}(t)
 + c_V\,(1-\gamma)\,\lambda_{Vi}(t)\,Nv_{i}(t)
 - Bv_{i}(t)\,(r_{Ni}+r_{Vi})
 + \pi_i(t)\,B_i(t) - \omega\,Bv_{i}(t).
\end{aligned}
\right\}
\end{equation}
The force of infection for both vaccine-type (VT) and non-vaccine-type (NVT) pneumococcal serotypes is age- and time-dependent, determined by the number of vaccinated and unvaccinated carriers in each age group $j$ and by the effective contact rate $\beta_{ij}$ between individuals in age group $i$ and those in group $j$, expressed as:
\begin{equation} 
\left.
    \begin{aligned}
\lambda_{Vi}(t) &= \sum_{j} \beta_{V,ij}\,
\bigl( V_j(t) + Vv_{j}(t) + B_j(t) + Bv_{j}(t) \bigr), \\[4pt]
\lambda_{Ni}(t) &= \sum_{j} \beta_{N,ij}\,
\bigl( N_j(t) + Nv_{j}(t) + B_j(t) + Bv_{j}(t) \bigr),
\end{aligned}
\right\}
\end{equation}
where
$$
\beta_{ij} = \varepsilon^{B}\,\beta^{a}_{ij} + (1-\varepsilon)\,\beta^{p}_{ij}, 0\leq \epsilon\leq 1, a=\text{assortative},~ p=\text{proportionate}. 
$$
The graphical representation of this model \eqref{vaccine_model} is shown in Figure \ref{fig:melegaro}.  The meanings of the variables and parameters   are shown in Figure \ref{fig:melegaro}.
This model is proposed for the bacterial infection caused by \textit{S. pneumoniae}  to predict the  impacts of  7-valent conjugate vaccine PCV7, on the incidence of invasive disease accounting for both herd immunity and serotype replacement effects.
The model \eqref{vaccine_model} consists of 100 cohorts of individuals (0, 1, 2, 3,..., 99) each corresponding to one year of age. The outcomes of this study \cite{2010_melegaro_dynamic} indicate that PCV7 vaccination programme could eliminate vaccine serotypes. However, the associated rise in carriage of non-vaccine serotypes and the potential increase in invasive disease could diminish, even outweigh the overall benefit. Choi \textit{et al.}  \cite{2017_choi} suggest that PCV7 vaccination may reduce overall invasive pneumococcal disease, especially among children, despite the rapid replacement of non-PCV7 serotypes. However,  the replacement of PCV7 with PCV13 would cause an over all decrease for  invasive pneumococcal disease incidence in all cases \cite{2012_choi_mathematical}. 

The Joint Committee on Vaccination and Immunization (JCVI) in  UK revised PCV13 vaccine recommendation in 2017, reducing the number of primary doses from 2+1 to 1+1. The updated 1+1 schedule now advises administration at 3 months and a booster dose at 12 months for infants \cite{2019_choi_estimated}. 
Recently, Chen \textit{et al.} \cite{2024_chen_re}  conducted a re-evaluation of the impacts and cost-effectiveness of pneumococcal conjugate vaccine introduction across 112 low- and middle-income countries. Their findings indicate that countries with delayed implementation of PCV programs or insufficient vaccination coverage have experienced significantly higher mortality rates from vaccine-preventable pneumococcal diseases. These findings highlight the critical need for the rapid scaling up of PCV programs to ensure high coverage and to maximize the public health impacts of the vaccine. Considering eight age cohorts and seven serotype groups aligned with the compositions of various pneumococcal vaccines, Horn \textit{et al.} \cite{2023_horn_mathematical} developed a  transmission model  with the objective of predicting carriage prevalence and the burden of invasive pneumococcal disease (IPD) for serotypes covered by both the vaccines PCV15 and PCV20. As a result, Horn \textit{et al.} \cite{2023_horn_mathematical} concluded that implementing next-generation PCVs in adults could play some crucial roles in reducing the rising burden of adult IPD. They also pointed out that PCV20 may be a suitable candidate, offering the broadest protection against pneumococcal disease. In a recent study, using a static model for the Netherlands, Boer \textit{et al.} \cite{2024_de_higher}  showed that for long-term cost-effectiveness, pneumococcal vaccination for older adults should include invasive serotypes not covered by the childhood immunization program.

The most recently approved vaccine is PCV21, known as CAPVAXIVE, which received U.S. FDA approval in June, 2024 for adults aged 18 and older \cite{merck_pneumonia,2025_campos_practice}. A single dose of PCV21 is recommended for all adults aged 65 and older. PCV21 is also recommended for adults aged 19–64 with certain risk factors, if they have not previously received any PCV or if their vaccination history is unknown. CAPVAXIVE is specifically designed to protect against 21 serotypes of \textit{S. pneumoniae}, covering approximately 84\% of invasive pneumococcal disease in adults 50 years and older, and has shown robust immune responses in both vaccine-naïve and vaccine-experienced adults during Phase 3 clinical trials \cite{merck_pneumonia}. CAPVAXIVE has also been approved in Canada and Australia, with regulatory reviews ongoing in other countries.
However, to the view point of overall effectiveness, public acceptance and cost, PCV20 (Prevnar20) has been considered as one of the  dominant vaccines for infant, as it  shows significant reductions in the healthcare costs compared to the other vaccines, such as PCV13, PCV15 and disease burden \cite{2024_rozenbaum_cost}.

On the other hand, the vaccine development for RSV infection remains a critical research priority due to its high morbidity and mortality among infants, immunocompromised populations and older adults. In this regard, it is important to note that the early vaccine efforts in the 1960s utilizing  the formalin-inactivated RSV led to  vaccine-enhanced disease, thereby delaying in the progress for several decades \cite{kim1969respiratory}. However,  advances  in  the structural biology, in particular the elucidation of the pre-fusion conformation of the F glycoprotein, have significantly transformed  the design of the vaccines \cite{mclellan2013structure}. 
This has resulted the development of stabilized pre-F protein-based vaccines that elicit strong neutralizing antibody responses \cite{graham2017vaccine}. Several vaccine platforms including mRNA, vector-based, and protein subunit vaccines have demonstrated promising results in clinical trials \cite{clark2024recent}. Recently, two vaccines, Arexvy (GSK) and Abrysvo (Pfizer), have received regulatory approval for use in older adults and maternal immunization, marking a major milestone in RSV prevention \cite{kampmann2023bivalent}. More recently, using data from New Zealand, Prasad \textit{et al.} \cite{prasad2021modelling} showed that a seasonal monoclonal antibody strategy yields a greater impact on RSV disease prevention than maternal vaccination.
\color{black}

\section{Discussions} \label{conclusion}
In this review, we  provide a comprehensive synthesis of the epidemiology and modeling of pneumonia caused by \textit{Streptococcus pneumoniae} and respiratory syncytial virus (RSV), with particular emphasis on the role of vaccination and statistical inference. We begin by outlining the global burden of pneumonia, highlighting the striking age-specific mortality patterns that affect children under five and older adults to a greater extent, and summarizing the clinical and biological features of pneumococcal and RSV infections. We then describe the development and implementation of pneumococcal conjugate vaccines (PCVs) and RSV vaccines or monoclonal antibodies, as well as their coverage, limitations, and serotype or strain-specific issues. 
Based on the inclusion–exclusion criteria and multiple reliable bibliographic databases, a wide spectrum of modeling frameworks is reviewed, beginning from the classical deterministic epidemic models (\textit{e.g.}, SIR and its extensions) to age-structured, vaccination models, seasonally forced models by incorporating  the waning immunity, the latency, carriage, serotype replacement and reinfection. In parallel, we also examine stochastic and state-space models, focusing how they taken into account  the demographic stochasticity, noisy surveillance data and under-ascertainment . In particular, we pay our attention  to the Bayesian workflow that links mechanistic models to observed data through posterior inference, principled prior specification, and model checking. This  enables us to estimate the key quantities, such as forces of infection, time-varying reproduction numbers. Finally, we  give a through discussion on vaccination models for pneumococcus and RSV infection that incorporate serotype groups, age structure and cost-effectiveness analysis to guide public health decision-making and  evaluate competing immunization strategies.

Note that our analysis underscores that vaccination is  the most powerful and widely applied intervention; that reduce the burden of both pneumococcal and RSV pneumonia. It is worth nothing that the widespread use of these PCVs resulted in substantial reduction in the invasive pneumococcal disease and pneumonia in vaccinated individuals, with add-on herd protection observed in unvaccinated cohorts. Preciously speaking, the successive vaccine inventions with taking broader range of serotypes (\textit{e.g.}, PCV7, PCV10, PCV13, PCV15, PCV20, and PCV21) \cite{2019_choi_estimated,2016_daniels,2011_choi} have further expanded protection. However,  and inequities in access and the concerns about serotype replacement still persist. Simultaneously, the recent advances in long-acting monoclonal antibodies and the development in RSV vaccine  have opened new avenues for protecting the high-risk infants, young children, immunocompromised groups and older adults. Recently, it is seen that seasonal monoclonal antibody strategy yields a greater impact on RSV disease prevention than maternal vaccination. Dynamic transmission models consistently demonstrate that vaccination can reduce basic and effective reproduction numbers, flatten epidemic peaks, and mitigate seasonal surges that strain healthcare systems. When combined with economic evaluation, these models show that appropriately targeted vaccination especially in high-burden, resource-limited settings can be cost-effective or even cost-saving, reducing hospitalizations, mortality, and long-term complications. Furthermore, the integration of vaccination strategies with broader public health measures, such as early detection, improved healthcare infrastructure, and effective treatment regimens, has the potential to further reduce the global burden of pneumonia and enhance the resilience of healthcare systems in responding to respiratory infectious.

At the same time, this review also highlights few important future directions that are very critical to refine the model-based evidence and optimize the pneumonia control. It is the proper place to mention that there is a need for an integrated  model that can capture pneumococcal–RSV co-infections dynamics, and the interactions with other respiratory pathogens, such as COVID-19. In addition to this, the impacts of changing demography, the climate variability on transmission dynamics, and  the contact patterns  should also be taken into account.  We  also emphasize that the multi-scale modeling framework that link the  within-host immune responses, pathogen, and population-level spread could play an important role in  improving the understanding of  vaccine escape mechanism, waning immunity, and the most importantly, serotype or strain replacement under different vaccination policies.

From a statistical standpoint, broader use of Bayesian state-space and hierarchical models fitted jointly to hospitalization, carriage, mortality, serological, and genomic data will be essential for robust inference and real-time forecasting, particularly in low- and middle-income countries where data are sparse or incomplete. Future work should also systematically compare next-generation PCVs and RSV products, explore optimal age-specific and maternal immunization schedules, and explicitly embed model outputs into policy processes. Closer collaboration between modelers, statisticians, clinicians, immunologists, and public health authorities will be crucial for translating these methodological advances into concrete, evidence-based strategies that reduce the global burden of pneumonia caused by \textit{S. pneumoniae} and RSV.

\subsection*{Acknowledgments}
\textcolor{black}{Funding for this study was provided by the Bill \& Melinda Gates Foundation, United States (NV-044445). Subhra Sankar Dhar also gratefully acknowledges his core research grant CRG/2022/001489, Government of India .}
\subsection*{Data availability statement}
 Data sharing is not applicable. 
 \subsection*{Author contributions}
 All the authors contribute equally.
 \subsection*{Conflict of interest}
The authors declare no potential conflict of interests.

\bibliography{Ref}
\end{document}